\documentclass[preprint]{aastex}

\shorttitle{VARIABLE IR COUNTERPART TO SGR 1806--20}
\shortauthors{KOSUGI, OGASAWARA, \& TERADA}

\begin{document}

\title{A VARIABLE INFRARED COUNTERPART TO THE SOFT GAMMA-RAY REPEATER SGR 1806--20\thanks{Based on data collected at the Subaru Telescope, which is operated by the National Astronomical Observatory of Japan.}}

\author{\sc{George Kosugi, Ryusuke Ogasawara, and Hiroshi Terada}}
\affil{Subaru Telescope, National Astronomical Observatory of Japan,
    650 N'Aohoku Place, Hilo, HI 96720}
\email{george@subaru.naoj.org, ryu@subaru.naoj.org, terada@subaru.naoj.org}

\begin{abstract}

We present $K\arcmin$-band (2.12 $\micron$) imaging observations of the SGR 1806--20 field taken during its very active phase in mid 2004, which reveal brightening of sources within the {\it{Chandra}} X-ray error circle when compared with earlier images obtained in 2002.
One source brightened by more than a factor of 2, and so we consider this to be the probable infrared counterpart for SGR 1806--20.
The other two sources are located in close proximity to the probable counterpart and show marginal brightening, which may suggest that the high-energy photons emitted from the SGR during its active phase have induced dust sublimation or brightening of the unresolved background around the SGR.

\end{abstract}


\section{INTRODUCTION}

The soft gamma-ray repeaters (SGRs) are now commonly believed to be magnetars, isolated young neutron stars with ultrahigh magnetic fields of $B > 10^{14}$ G \citep[e.g.,][]{dun1992}.
Detections of long period X-ray pulsations and their rapid spin-down rates \citep[$P$ = 5--8 s and $\dot{P} \sim 10^{-11}$ ss$^{-1}$, respectively; e.g.,][]{kou1998,kou1999,hur1999}, cyclotron resonance features in the burst X-ray spectrum \citep{ibr2002}, and burst energetics \citep[e.g.,][]{pac1992,tho1995} all strongly support the magnetar model.

Another small class of objects, the anomalous X-ray pulsars \citep[AXPs:][]{mer1995}, are also presumed to be magnetars, owing to their similarities with SGRs both in spin period and deceleration rate.
In addition, the discovery of several SGR-like X-ray bursts from AXPs \citep{gav2002,kas2003} has increased the evidence for a connection between SGRs and AXPs.
To date, possible optical-infrared (optical-IR) counterparts with unusual colors have been found for five out of six AXPs.
Four show variability in some way correlated with changes in
X-ray flux (\citealp[4U 0142+61:][]{ker2002,hul2004}; \citealp[1E 2259+586:][]{kas2003}; \citealp[1E 1048.1-5937:][]{isr2002}; \citealp[and XTE J1810-19704:][]{rea2004}).
In the case of AXP 1E 2259+586, the $K_{s}$-band flux increased by a factor of $\sim3$ by the third day after an X-ray burst.
Thus, variability is thought to be a common characteristic of the optical-IR counterparts for AXPs.
In contrast, none of the five SGRs have convincing counterparts detected with optical-IR variability, despite considerable observational effort.
The detection of optical-IR counterparts for SGRs is essential in order to investigate the connection between SGRs and AXPs.
If optical-IR counterparts for SGRs show a relation between
variability and burst activity similar to the AXPs, it will enable much easier detection of SGR counterparts after outbursts.

SGR 1806--20 gradually entered an active phase in late 2003
\citep[e.g.,][]{hur2003}.
It showed recurrent intense bursts in mid 2004, and a series of many bursts in late 2004.
Finally, a giant flare, the first observed from SGR 1806--20, occurred on 2004 December 27 \citep[e.g.,][]{bor2004,hur2005}, and left a fading radio afterglow with a precise localization \citep[e.g.,][]{cam2005,gae2005}.
A series of infrared observations were performed with the Subaru 8.2-m telescope during the active phase of SGR 1806--20 in mid -- late 2004, several months prior to the giant flare.
Here we report the discovery of the probable infrared counterpart for SGR 1806--20 within the {\it{Chandra}} X-ray error circle.
We also discuss the influence of the high-energy photons emitted by the SGR burst upon the surroundings.

\section{OBSERVATIONS, DATA ANALYSIS, AND RESULTS}

\subsection{Infrared Data and Photometry}

Our first epoch observations were obtained on 2002 May 25 in $J$ (1.25 $\micron$), $H$ (1.63 $\micron$), and $K\arcmin$ (2.12 $\micron$) bands using the infrared camera and spectrograph \citep[IRCS:][]{kob2000} on the Subaru 8.2-m telescope \citep{iye2004} under photometric sky conditions.
A plate scale of $0\farcs058$ pixel$^{-1}$ was chosen so as to match the average seeing conditions (FWHM = $0\farcs30$ in $K\arcmin$-band) on the night of the observation.
The field of view (FOV) was approximately $1\arcmin \times 1\arcmin$.
Images were taken by using nine-point dithering of three coadds with 30-s exposures each, and the resulting total integration time was 810 s for all three bands.
A UKIRT faint standard star FS 148 \citep{cas1992} was also observed using the same configuration.
We carried out a second epoch of observations in the $K\arcmin$ band on 2004 August 8, during the active
phase, with the same configurations as for the first epoch, to allow a precise comparison of the images from the two epochs.
The sky conditions were again excellent, and the seeing size was $0\farcs32$.
A total integration time of 1260 s was achieved with a nine- and a five-point dithering of 90-s exposures at each position.
We also performed another observation in the $K\arcmin$ band with adaptive optics \citep[AO:][]{tak2004} on the same night using a 13.7-th magnitude star located 30" north of the SGR as the natural reference star.
The plate scale was set to $0\farcs023$ pixel$^{-1}$ (FOV was $23\arcsec $ square), and the stellar image size improved to $0\farcs20$.
Two sets of nine-point dithering observations of 90-s exposures at each position were obtained.
Fourteen frames with good seeing were selected, and the resulting total integration time was 1260 s.
The 3rd and the 4th epoch observations were carried out using AO on 2004 September 3 and 23, respectively, also during the active phase.
The best quality frames with three coadds of 30-s exposures were selected to create final images for those epochs.
The resulting total integration time and the stellar image size were 1710 s and $0\farcs22$ for the 3rd epoch, and 1980 s and $0\farcs38$ for the 4th epoch.

The standard imaging calibration procedures were applied for each epoch data using the IRAF data analysis software.\footnote{IRAF is distributed by the National Optical Astronomy Observatories, which are operated by the Association of Universities for Research in Astronomy, Inc., under cooperative agreement with the National Science Foundation.}
There is some latency in the ALADDIN array when bright objects are focused on it; therefore, the locations of bright objects in the previous frame were masked before stacking selected images.
Then, a photometry of the field stars in the vicinity of the SGR X-ray position was carried out with the  DAOPHOT/ALLSTAR package \citep{ste1987} in IRAF.
The radius of the point spread function (PSF) was set to twice the stellar FWHM at each epoch, and the aperture radius for the photometry was set equal to the FWHM.
We applied the photometric standard star FS 148 to our first epoch data and measured a field star C \citep[see Figure 3 in] [and Figure 2D in this paper]{eik2001} as a secondary standard.
The $K\arcmin$-band magnitude of star C was $K\arcmin$ = 16.48 $\pm$ 0.05, which is consistent with previous work by \citet{wac2004} but slightly different to that of \citet{eik2001}.
For the other epoch data, photometry was conducted relative to star C, based on the assumption that the star is not variable.
The 3-$\sigma$ limiting magnitudes were 21.5, 21.4, 21.9, 22.0, and 21.0 for the 1st epoch, 2nd epoch without AO, 2nd epoch with AO, 3rd epoch, and 4th epoch data, respectively.

\subsection{Optical Data and Astrometry}

An unfiltered optical image of a 10-s exposure, covering a FOV of $6\arcmin$ diameter, was taken with the faint optical camera and spectrograph \citep[FOCAS:][]{kas2002} on the Subaru telescope on 2002 June 18.
As the seeing was $0\farcs9$ and the limiting magnitude was sufficiently deep, we used this image as a reference to obtain precise astrometry for our IR data.
First, the positions of 22 unsaturated, stellar-like sources in the optical image were measured and compared with the USNO-B1.0 catalog \citep{mon2003}.
The optical image was then compared with the $J$-band image using 20 stars common to both images.
Finally, the $J$-band image was compared with the $K\arcmin$-band image using another 16 stars.
Taking the positional errors of the individual stars from the USNO-B1.0 catalog into account, the
astrometric errors incurred in these procedures were $0\farcs06$, $0\farcs01$, and less than $0\farcs01$, respectively, in each coordinate.
An empirical value for the size of local offsets in the southern part of the USNO-B1.0 catalog compared to the International Celestial Reference Frame (ICRF) is $0\farcs136$ \citep{sil2005}.
The resulting final astrometric uncertainties in our $K\arcmin$-band images are $0\farcs15$.
Thus, the 99\% uncertainties for the {\it{Chandra}} X-ray error radius \citep[$0\farcs3$;][]{kap2002} and the VLA radio afterglow error radius \citep[$0\farcs1$;][]{cam2005} on our infrared images are $0\farcs77$ and $0\farcs41$, respectively.
The positions of the field stars, along with the 2nd epoch photometry, are summarized in Table \ref{tbl-2}.

\subsection{Field Around SGR 1806--20}

The SGR 1806--20 field along with its putative associated massive cluster is shown in Figure 1A.
A luminous blue variable star LBV 1806--20 is seen $12\arcsec$ east of the center.
Bright cluster stars are located mainly to the north to north-west of the SGR position.
As mentioned by \citet{cor2004}, the color of these stars indicates that they suffer large extinction ($A_{V} \sim$ 30).
A region of even higher reddening can be observed about $25\arcsec$ to the north-west of the center.
The close-up (Figure 1B) shows a lot of faint and redder, consequently probable cluster member, stars.
There are two remarkable moniliform stellar arcs $2\farcs1$ north and $3\farcs6$ south-west of the SGR.  Unfortunately, ghost images of the bright star, produced by the compensator inside the IRCS optics, overlap with part of the stellar arc to the south-west.
It is intriguing that the center of curvature of both arcs is located close to the SGR position.

Multi-epoch $K\arcmin$-band images are presented in Figure 2.
Figure 2D shows the {\it{Chandra}} and VLA positions with 99\% confidence error circles (see 2.2) along with the labels of individual field stars.
Stars A, B, and C in \citet{eik2001} correspond to our stars A1 to A4, B1 to B3, and C, respectively.
Better spatial resolution and a deeper detection limit enabled us to spot the multiplicity of the stars A and B.
A remarkable difference can be seen between Figure 2A and 2B: star B3 can hardly be seen in the 1st epoch but is bright in the 2nd epoch.
There is no other major difference between the two images.
All good quality images taken with AO are accumulated to create a high-resolution image (Figure 2C; FWHM = $0\farcs18$).

The light curve of each field star is presented in Figure 3.
Most of the stars are constant within the error bars, which implies that star C, used in calibration, was constant during the period covered by our observations.
B3 shows brightening by more than a factor of 2, whereas A3 shows a small amount of fading.
B1 and B2 show slight brightening that just exceeds the error bars.

\section{DISCUSSIONS AND CONCLUSIONS}

\subsection{Detection of a Variable IR Counterpart}

The position of SGR 1806--20 is well localized in the X-ray; however, the field of SGR 1806--20 is very crowded at IR wavelengths and the probability of a chance coincidence with unrelated IR sources is relatively high.
Therefore, variability is a key means of distinguishing the real IR counterpart.
The IR source B3 is the only source to show brightening in the $K\arcmin$ band clearly during the active phase.
The chance probability of finding a variable star is generally low, i.e., $4.1 \times 10^{-3}$ is derived from the Optical Gravitational Lensing Experiment (OGLE-II) dataset \citep{zeb2001}, and is much lower if we select highly variable stars like B3.

Recently, the persistent hard X-ray emission ($\lesssim$200 keV) from SGR 1806--20 was presented, and its spectral shape was different from that of the bright bursts \citep{mol2004}.
It is also to be noted that the persistent hard X-ray emission correlates in intensity and spectral hardness with the level of bursting activity \citep{mer2004}.
The flux in the 20--100 keV energy band was almost doubled during the active phase in 2004 September to October compared with the data in 2003.
On the other hands, our first epoch data was taken in 2002, and the IR flux for B3 increased by more than a factor of 2 in our third epoch in 2004.
It is likely that the IR flux would increase with the persistent X-ray flux.

We should also note here that two short bursts were detected on 2002 August 25 by Konus-Wind and Ulysses \citep{hur2002}, just 1.5 and 9 hours prior to our $K\arcmin$-band observations.
Our second epoch observations were made 11.4 and 15.9 days after the short bursts on July 28 and July 23 with fluences of $4.4 \times 10^{-6}$ erg cm$^{-2}$ and $7.5 \times 10^{-6}$ erg cm$^{-2}$, respectively \citep{gol2004a}.
Our third epoch data were collected 5.8 and 8.3 days after two intense intermediate bursts on August 28 ($4.0 \times 10^{-5}$ erg cm$^{-2}$; \citealp{gol2004c}) and August 25 ($2.1 \times 10^{-5}$ erg cm$^{-2}$; \citealp{gol2004b}), respectively.
Some intense SGR bursts are followed by an X-ray tail or afterglow that decays with time \citep[e.g.,][]{ibr2001, len2003}.
Assuming that the SGR has an IR afterglow associated with X-ray burst activity, by analogy with AXPs (e.g., \citealp[AXP 1E 2259+586:][]{kas2003,tam2004}; and \citealp[XTE J1810-197:][]{rea2004}), SGR 1806--20's IR counterpart would be expected to be brighter at the third epoch than at the second, because the elapsed time since the last burst was shorter and the fluence of the burst was larger.
The light curve of B3 between the 2nd and 3rd epochs is also consistent with this model.
On the other hand, the elapsed time period since the last burst for the 1st epoch data is much shorter than for the 2nd epoch data, but B3 is only marginally detected in the 1st epoch.
This implies either that 1) time lags exist between X-ray and IR flux enhancements, or 2) multiple burst activities are in some way accumulated in the IR.

An alternative interpretation of the brightening behavior of B3 is that B3 could have a large proper motion ($0\farcs1$ yr$^{-1}$) and so was very close to B1 in the 1st epoch.
In this scenario the difficulties in accurately separating B1 and B3 may have caused the apparent brightening of B1 and fading of B3 at the 1st epoch.
Future high-resolution imaging observations are necessary in order to measure precisely the proper motion of these sources.

\subsection{Brightening of Nearby Stars}

The active phase of SGR 1806--20 has been ongoing for several months, and so the inter-stellar medium (ISM) surrounding the SGR may be affected by the presence of high energy photons.
In the case of gamma-ray bursts, for example, a large amount of dust sublimation is expected by the optical-UV prompt emission \citep{wax2000} and X-ray radiation \citep{fru2001}.
It is suspected that SGR 1806--20 is associated with a massive molecular cloud \citep{cor2004}, and suffers a large extinction ($A_{V} >$ 30) inferred from its X-ray column density of $N_{H} \approx 6 \times 10^{22}$ cm$^{-2}$ \citep{son1994,mer2000} and the relation of $A_{V}/N_{H} = 5.6 \times 10^{-22}$ \citep{pre1995}.
As the burst activity has been ongoing for a considerable period, the dust in a spherical region centered on the SGR is expected to have been sublimated, and hence the brightness of the background stars close to the line of sight of SGR 1806--20 might be expected to brighten.
In fact, we marginally detected subtle brightening of the nearby stars B1 and B2.
However, the brightening of the nearby stars could also be interpreted as the brightening of unresolved
background (nebulosity) owing to the heating of the dust surrounding the SGR.
We can discriminate between these two mechanisms by measuring the variability in different wavelengths.
In the first scenario, the color of the stars may change, i.e. they will be brighter at shorter wavelengths, and the scale of the variability follows the extinction law.
In the latter case, the variability may be larger at longer wavelengths.
The distances of the stars B1 and B2 from the line of sight toward B3 are $0\farcs23$ and $0\farcs30$, which correspond to 20.3 and 26.0 light-days, respectively, at a distance of 15.1 kpc \citep{cor2004}.
Therefore, the burst photons could clearly have reached and affected the ISM along the line of sight towards B1 and B2 when the observations were obtained.
Further multicolor monitoring is required to confirm these conclusions.

\acknowledgments

We would like to thank all staff members of the Subaru Telescope for their observational and data analysis support, and the FOCAS team members for the use of the data taken in their scientific verification phase.
We also express our gratitude to the referee Kevin Hurley for his helpful comments and suggestions.

%We also thank XXXX for his helpful comments.

%Facilities: \facility{Subaru Telescope}.

\clearpage

\begin{figure}
\epsscale{1.00}
%\plotone{f1.eps}
\caption{
Wide-field ($55 \arcsec \times 55 \arcsec$) $J$, $H$, and $K\arcmin$-band three-color composite image of the SGR 1806--20 field taken on 2002 May 25 ($\it{left}$).
Blue, green, and red correspond to the $J$, $H$, and $K\arcmin$ bands, respectively.
North is up and East is left.
The elongated sources at $16\farcs9$ north and west from bright stars are ghost images caused by the beamsplitter and compensator, respectively.
A close-up of the central region (gray square with $10 \arcsec \times 10 \arcsec$) is shown in the $\it{right}$ panel.
A gray circle represents the {\it{Chandra}} X-ray error circle \citep{kap2002} with a radius of $0\farcs77$ (see 2.2).
A compensator ghost is seen to the south-west of center.
\label{fig1}}
\end{figure}

\clearpage

\begin{figure}
\epsscale{1.00}
%\plotone{f2.eps}
\caption{
(A) $K\arcmin$-band image taken on 2002 May 25 (1st epoch) without using AO.
(B) $K\arcmin$-band image taken on 2004 August 8 (2nd epoch) without AO.
(C) $K\arcmin$-band image taken with AO.
Only images with smaller FWHM in the 2nd and the 3rd epoch are combined.
The final stellar size is $0\farcs18$ in FWHM.
(D) Same image as (C) is shown with a different intensity scale, showing the $0\farcs77$ radius error circle around the {\it{Chandra}} X-ray position of $\alpha$ = $18^{h}08^{m}39^{s}.32$ and $\delta$ = $-20^{\circ}24\arcmin39\farcs5$ (J2000.0), the $0\farcs41$ radius error circle around the VLA radio afterglow position of $\alpha$ = $18^{h}08^{m}39^{s}.34$ and $\delta$ = $-20^{\circ}24\arcmin39\farcs7$ (J2000.0), and the star labels in Table 1.
\label{fig2}}
\end{figure}

\clearpage

\begin{figure}
\epsscale{0.8}
\plotone{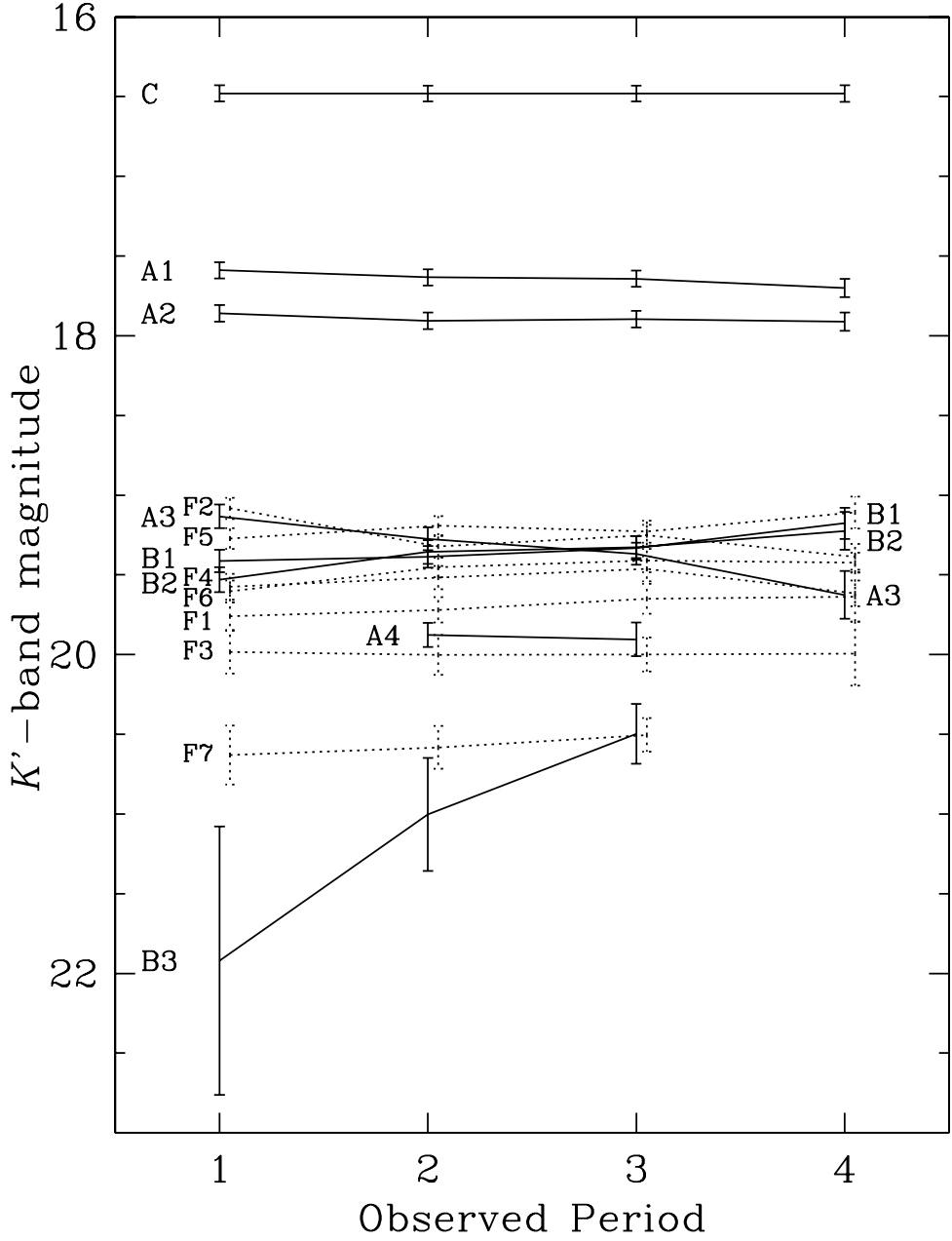}
\caption{
$K\arcmin$-band light curves of the SGR 1806--20 field stars are presented.
Solid lines are used for the stars A1 -- A4, B1 -- B3, and C, and dashed lines for the stars F1 -- F7.
Owing to the larger plate scale, we were unable to separate star A4 from A1 in the $0\farcs058$ pixel-scale images (without AO).
Therefore, the flux from A4 contaminates the flux of A1 in the 1st epoch, and hence the magnitude of A1 may be overestimated by about 0.1 mag.
B3, F7 and A4 were not measured in the 4th epoch data owing to the relatively poor seeing conditions.
The 2nd epoch plots are based on the observation with AO.
\label{fig3}}
\end{figure}

\clearpage

\begin{deluxetable}{ccccc}

\tablecaption{\sc{Astrometry and Photometry of the Point Sources Near SGR 1806--20}\label{tbl-2}}

\tablenum{1}

\tablehead{\colhead{Label} & \colhead{RA(J2000)} & \colhead{DEC(J2000)} & \colhead{$K\arcmin$ magnitude}} 

\startdata
A1 & 18 08 39.386 & -20 24 39.02 & 17.63 $\pm$ 0.05 \\
A2 & 18 08 39.369 & -20 24 39.43 & 17.91 $\pm$ 0.05 \\
A3 & 18 08 39.358 & -20 24 39.01 & 19.27 $\pm$ 0.07 \\
A4 & 18 08 39.383 & -20 24 39.18 & 19.88 $\pm$ 0.08 \\
B1 & 18 08 39.340 & -20 24 39.77 & 19.39 $\pm$ 0.07 \\
B2 & 18 08 39.343 & -20 24 40.14 & 19.36 $\pm$ 0.07 \\
B3 & 18 08 39.329 & -20 24 39.94 & 21.00 $\pm$ 0.36 \\
C  & 18 08 39.319 & -20 24 40.81 & 16.48 $\pm$ 0.05 \\
F1 & 18 08 39.382 & -20 24 40.34 & 19.72 $\pm$ 0.08 \\
F2 & 18 08 39.290 & -20 24 40.31 & 19.32 $\pm$ 0.07 \\
F3 & 18 08 39.277 & -20 24 39.26 & 20.00 $\pm$ 0.12 \\
F4 & 18 08 39.257 & -20 24 39.24 & 19.52 $\pm$ 0.07 \\
F5 & 18 08 39.214 & -20 24 40.27 & 19.19 $\pm$ 0.06 \\
F6 & 18 08 39.229 & -20 24 40.85 & 19.45 $\pm$ 0.06 \\
F7 & 18 08 39.223 & -20 24 38.63 & 20.58 $\pm$ 0.13 \\
\enddata

\tablecomments{Measurements are based on the dataset taken with AO on 2004 August 8.
Units of right ascension are hours, minutes, and seconds, and units of declination are degrees, arcminutes, and arcseconds.
Astrometric uncertainties are $0\farcs15$ in each coordinate.}

\end{deluxetable}
\end{document}